\documentclass[pdflatex,sn-aps]{sn-jnl}

\jyear{2021}%

\theoremstyle{thmstyleone}%
\theoremstyle{thmstyletwo}%
\theoremstyle{thmstylethree}%

\newcommand{\daksha}{\textit{Daksha}}
\newcommand{\asat}{{\em AstroSat}}

\newcommand{\dakshasci}{Paper~II}

\newcommand{\degr}{\ensuremath{^\circ}}

\newcommand{\swift}{\emph{Swift}}
\newcommand{\bat}{\swift-BAT}
\newcommand{\fermi}{\emph{Fermi}}
\newcommand{\gbm}{\fermi-GBM}
\newcommand{\svom}{\emph{SVOM}}
\newcommand{\eclairs}{\svom/ECLAIRs}
\newcommand{\theseus}{\emph{THESEUS}}
\newcommand{\xgis}{\theseus/XGIS}

\newcommand{\ecs}{\ensuremath{\mathrm{erg~cm}^{-2}~\mathrm{s}^{-1}}}
\newcommand{\pcs}{\ensuremath{\mathrm{ph~cm}^{-2}~\mathrm{s}^{-1}}}
\newcommand{\ee}[1]{\ensuremath{\times 10^{#1}}}
\newcommand{\eu}[2]{\ensuremath{\times 10^{#1}~\mathrm{#2}}}

\makeatletter
\let\jnl@style=\rm
\def\ref@jnl#1{{\jnl@style#1}}

\def\aj{\ref@jnl{AJ}}                   
\def\actaa{\ref@jnl{Acta Astron.}}      
\def\araa{\ref@jnl{ARA\&A}}             
\def\apj{\ref@jnl{ApJ}}                 
\def\apjl{\ref@jnl{ApJ}}                
\def\apjs{\ref@jnl{ApJS}}               
\def\ao{\ref@jnl{Appl.~Opt.}}           
\def\apss{\ref@jnl{Ap\&SS}}             
\def\aap{\ref@jnl{A\&A}}                
\def\aapr{\ref@jnl{A\&A~Rev.}}          
\def\aaps{\ref@jnl{A\&AS}}              
\def\azh{\ref@jnl{AZh}}                 
\def\baas{\ref@jnl{BAAS}}               
\def\bac{\ref@jnl{Bull. astr. Inst. Czechosl.}}
\def\caa{\ref@jnl{Chinese Astron. Astrophys.}}
\def\cjaa{\ref@jnl{Chinese J. Astron. Astrophys.}}
\def\icarus{\ref@jnl{Icarus}}           
\def\jcap{\ref@jnl{J. Cosmology Astropart. Phys.}}
\def\jrasc{\ref@jnl{JRASC}}             
\def\memras{\ref@jnl{MmRAS}}            
\def\mnras{\ref@jnl{MNRAS}}             
\def\na{\ref@jnl{New A}}                
\def\nar{\ref@jnl{New A Rev.}}          
\def\pra{\ref@jnl{Phys.~Rev.~A}}        
\def\prb{\ref@jnl{Phys.~Rev.~B}}        
\def\prc{\ref@jnl{Phys.~Rev.~C}}        
\def\prd{\ref@jnl{Phys.~Rev.~D}}        
\def\pre{\ref@jnl{Phys.~Rev.~E}}        
\def\prl{\ref@jnl{Phys.~Rev.~Lett.}}    
\def\pasa{\ref@jnl{PASA}}               
\def\pasp{\ref@jnl{PASP}}               
\def\pasj{\ref@jnl{PASJ}}               
\def\rmxaa{\ref@jnl{Rev. Mexicana Astron. Astrofis.}}%
\def\qjras{\ref@jnl{QJRAS}}             
\def\skytel{\ref@jnl{S\&T}}             
\def\solphys{\ref@jnl{Sol.~Phys.}}      
\def\sovast{\ref@jnl{Soviet~Ast.}}      
\def\ssr{\ref@jnl{Space~Sci.~Rev.}}     
\def\zap{\ref@jnl{ZAp}}                 
\def\nat{\ref@jnl{Nature}}              
\def\iaucirc{\ref@jnl{IAU~Circ.}}       
\def\aplett{\ref@jnl{Astrophys.~Lett.}} 
\def\apspr{\ref@jnl{Astrophys.~Space~Phys.~Res.}}
\def\bain{\ref@jnl{Bull.~Astron.~Inst.~Netherlands}} 
\def\fcp{\ref@jnl{Fund.~Cosmic~Phys.}}  
\def\gca{\ref@jnl{Geochim.~Cosmochim.~Acta}}   
\def\grl{\ref@jnl{Geophys.~Res.~Lett.}} 
\def\jcp{\ref@jnl{J.~Chem.~Phys.}}      
\def\jgr{\ref@jnl{J.~Geophys.~Res.}}    
\def\jqsrt{\ref@jnl{J.~Quant.~Spec.~Radiat.~Transf.}}
\def\memsai{\ref@jnl{Mem.~Soc.~Astron.~Italiana}}
\def\nphysa{\ref@jnl{Nucl.~Phys.~A}}   
\def\physrep{\ref@jnl{Phys.~Rep.}}   
\def\physscr{\ref@jnl{Phys.~Scr}}   
\def\planss{\ref@jnl{Planet.~Space~Sci.}}   
\def\procspie{\ref@jnl{Proc.~SPIE}}   

\makeatother

\usepackage{multirow}
\usepackage{multicol}
\newcommand{\orcid}[1]{\href{https://orcid.org/#1}{\includegraphics[scale=.05]{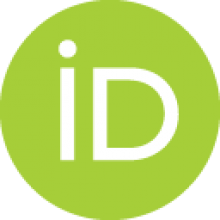}}}

\begin{document}

\title[Daksha Instrument]{Daksha: On Alert for High Energy Transients}


\author*[1]{\fnm{Varun} \sur{Bhalerao}\orcid{0000-0002-6112-7609} } \email{varunb@iitb.ac.in}
\author[2]{\fnm{Santosh} \sur{Vadawale}~\orcid{0000-0002-2050-0913}}
\author[3]{\fnm{Shriharsh} \sur{Tendulkar}~\orcid{0000-0003-2548-2926}}
\author[4]{\fnm{Dipankar} \sur{Bhattacharya}~\orcid{0000-0003-3352-3142}}
\author[5]{\fnm{Vikram} \sur{Rana}~\orcid{0000-0003-1703-8796}}

\author[2]{\fnm{Hitesh} \sur{Kumar L. Adalja}~\orcid{0000-0002-5272-6386}}
\author[6]{\fnm{Hrishikesh} \sur{Belatikar}~\orcid{0000-0001-9954-7329}}
\author[6]{\fnm{Mahesh} \sur{Bhaganagare}}
\author[7]{\fnm{Gulab} \sur{Dewangan}~\orcid{0000-0003-1589-2075}}
\author[1]{\fnm{Abhijeet} \sur{Ghodgaonkar}~\orcid{0000-0001-6211-2209}}
\author[2]{\fnm{Shiv} \sur{Kumar Goyal}~\orcid{0000-0002-3153-537X}}
\author[8]{\fnm{Suresh} \sur{Gunasekaran}}
\author[8]{\fnm{Guruprasad} \sur{P J}~\orcid{0000-0003-0162-0132}}
\author[3]{\fnm{Jayprakash} \sur{G. Koyande}~\orcid{0000-0001-7829-3366}}
\author[9]{\fnm{Salil} \sur{Kulkarni}}
\author[3]{\fnm{APK} \sur{Kutty}}
\author[2]{\fnm{Tinkal} \sur{Ladiya}~\orcid{0000-0001-6022-8283}}
\author[1]{\fnm{Suddhasatta} \sur{Mahapatra}}
\author[9]{\fnm{Deepak} \sur{Marla}~\orcid{0000-0001-6829-7830}}
\author[3]{\fnm{Sujay} \sur{Mate}~\orcid{0000-0001-5536-4635}}
\author[2]{\fnm{N.P.S.} \sur{Mithun}~\orcid{0000-0003-3431-6110}}
\author[9]{\fnm{Rakesh} \sur{Mote}~\orcid{0000-0001-7853-5150}}
\author[6]{\fnm{Sanjoli} \sur{Narang}~\orcid{0000-0002-8723-2263}}
\author[3]{\fnm{Ayush} \sur{Nema}~\orcid{0000-0003-0701-9639}}
\author[6]{\fnm{Sudhanshu} \sur{Nimbalkar}~\orcid{0000-0002-2131-6720}}
\author[1]{\fnm{Archana} \sur{Pai}~\orcid{0000-0003-3476-4589}}
\author[1]{\fnm{Sourav} \sur{Palit}~\orcid{0000-0003-2932-3666}}
\author[2]{\fnm{Arpit} \sur{Patel}~\orcid{0000-0002-0929-1401}}
\author[8,10]{\fnm{Jinaykumar} \sur{Patel}~\orcid{0000-0002-6551-0963}}
\author[11]{\fnm{Priya} \sur{Pradeep}}
\author[8]{\fnm{Prabhu} \sur{Ramachandran}~\orcid{0000-0001-6337-1720}}
\author[2]{\fnm{B.S.} \sur{Bharath Saiguhan}~\orcid{0000-0001-7580-364X}}
\author[1]{\fnm{Divita} \sur{Saraogi}~\orcid{0000-0001-6332-1723}}
\author[1]{\fnm{Disha} \sur{Sawant}~\orcid{0000-0002-9702-6324}}
\author[2]{\fnm{M.} \sur{Shanmugam}~\orcid{0000-0002-5995-8681}}
\author[2]{\fnm{Piyush} \sur{Sharma}~\orcid{0000-0001-9670-1511}}
\author[6]{\fnm{Amit} \sur{Shetye}}
\author[2]{\fnm{Nishant} \sur{Singh}~\orcid{0000-0002-1975-0552}}
\author[12]{\fnm{Shreeya} \sur{Singh}}
\author[1]{\fnm{Akshat} \sur{Singhal}~\orcid{0000-0003-1275-1904}}
\author[11]{\fnm{S.} \sur{Sreekumar}}
\author[8,13]{\fnm{Srividhya} \sur{Sridhar}~\orcid{0000-0002-7972-169X}}
\author[14,15,1]{\fnm{Rahul} \sur{Srinivasan}~\orcid{0000-0002-7176-6690}}
\author[6]{\fnm{Siddharth} \sur{Tallur}~\orcid{0000-0003-1399-2187}}
\author[2]{\fnm{Neeraj} \sur{K. Tiwari}~\orcid{0000-0003-4269-340X.}}
\author[8]{\fnm{Amrutha} \sur{Lakshmi Vadladi}~\orcid{0000-0003-0814-9064}}
\author[2]{\fnm{C.} \sur{S. Vaishnava}~\orcid{0000-0002-8096-5683}}
\author[3]{\fnm{Sandeep} \sur{Vishwakarma}~\orcid{0000-0001-8159-5656}}
\author[1]{\fnm{Gaurav} \sur{Waratkar}~\orcid{0000-0003-3630-9440}}

\affil*[1]{\orgdiv{Department of Physics}, \orgname{IIT Bombay}, \orgaddress{\street{Powai}, \city{Mumbai}, \postcode{400076}, \country{India}}}
\affil[2]{\orgname{Physical Research Laboratory}, \orgaddress{Navrangpura}, \city{Ahmedabad}, \postcode{380009}, \country{India}}
\affil[3]{\orgname{Department of Astronomy and Astrophysics}, \orgaddress{Tata Institute of Fundamental Research}, \city{Mumbai}, \postcode{400005}, \country{India}}
\affil[4]{\orgdiv{Ashoka University}, \orgname{Department of Physics}, \orgaddress{Sonepat}, \city{Haryana}, \postcode{131029}, \country{India}}
\affil[5]{\orgdiv{Raman Research Institute}, \orgname{C. V. Raman Avenue}, \orgaddress{Sadashivanagar}, \city{Bengaluru}, \postcode{560080}, \country{India}}

\affil[6]{\orgdiv{Department of Electrical Engineering}, \orgname{IIT Bombay}, \orgaddress{Powai}, \city{Mumbai}, \postcode{400076}, \country{India}}
\affil[7]{\orgname{Inter-University Center for Astronomy and Astrophysics}, \city{Pune}, \state{Maharashtra}, \postcode{411007}, \country{India}}
\affil[8]{\orgdiv{Department of Aerospace Engineering}, \orgname{IIT Bombay}, \orgaddress{Powai}, \city{Mumbai}, \postcode{400076}, \country{India}}
\affil[9]{\orgdiv{Department of Mechanical Engineering}, \orgname{IIT Bombay}, \orgaddress{Powai}, \city{Mumbai}, \postcode{400076}, \country{India}}
\affil[10]{\orgdiv{Department of Mechanical and Aerospace Engineering}, \orgname{The University of Texas at Arlington}, \orgaddress{Arlington}, \city{TX}, \postcode{76019}, \country{USA}}
\affil[11]{\orgname{Vikram Sarabhai Space Centre}, \orgaddress{Kochuveli}, \city{Thiruvananthapuram}, \postcode{695022}, \country{India}}
\affil[12]{\orgdiv{Department of Chemical Engineering}, \orgname{IIT Bombay}, \orgaddress{Powai}, \city{Mumbai}, \postcode{400076}, \country{India}}
\affil[13]{\orgdiv{Department of Aerospace}, \orgname{University of Illinois Urbana-Champaign}, \postcode{61801}, \country{US}}
\affil[14]{\orgdiv{Universit\'{e} C\^{o}te d'Azur, Observatoire de la C\^{o}te d'Azur}, \orgname{CNRS, Laboratoire Lagrange, Bd de l'Observatoire}, \orgaddress{CS, 34229, 06304 Nice cedex 4}, \country{France}}
\affil[15]{\orgdiv{Artemis, Universit\'{e} C\^{o}te d'Azur, Observatoire de la C\^{o}te d'Azur}, \orgname{CNRS}, \postcode{F-06304}, \city{Nice}, \country{France}}

\abstract{We present \daksha, a proposed high energy transients mission for the study of electromagnetic counterparts of gravitational wave sources, and gamma ray bursts. \daksha\ will comprise of two satellites in low earth equatorial orbits, on opposite sides of earth. Each satellite will carry three types of detectors to cover the entire sky in an energy range from 1~keV to $>1$~MeV. Any transients detected on-board will be announced publicly within minutes of discovery. All photon data will be downloaded in ground station passes to obtain source positions, spectra, and light curves. In addition, \daksha\ will address a wide range of science cases including monitoring X-ray pulsars, studies of magnetars, solar flares, searches for fast radio burst counterparts, routine monitoring of bright persistent high energy sources, terrestrial gamma-ray flashes, and probing primordial black hole abundances through lensing.
In this paper, we discuss the technical capabilities of \daksha, while the detailed science case is discussed in a separate paper.}

\keywords{space vehicles: instruments --- instrumentation: detectors}



\maketitle

\section{Introduction}\label{sec1}
\daksha\ is a proposed high-energy transients mission comprising of two satellites monitoring the entire sky in the 1---1000~keV range. The primary science goals of \daksha\ are the study of electromagnetic counterparts to gravitational wave sources, and prompt emission gamma ray bursts (GRBs). In addition, \daksha\ will also study X-ray pulsars, create a high-energy all-sky map with Compton imaging, and detect X-ray polarization from transients and pulsars. 

The key science goal driving the \daksha\ mission is the detection of high energy counterparts to gravitational wave events. Multi-messenger astrophysics has received a great boost in the last decade with the joint detection of gravitational waves (GW) and electromagnetic (EM) radiation from a binary neutron star merger, GW170817 \citep{LIGOScientificCollaboration2017,gw170817}. This was a treasure--trove of physics --- in addition to being the first direct proof of a link between short GRBs and binary neutron star (BNS) mergers, it provided insights into r-process nucleosynthesis \citep{whs+19,kkl+19}, yielded an independent measurement of $H_0$ \citep{TheLIGOScientificCollaboration2019}, measurements of equation of state \citep{De2018}. However, no EM counterparts could be found for  subsequent binary neutron star mergers detected by the advanced GW networks~\citep[see for instance][]{Hosseinzadeh2019,Coughlin2019}. This was not very surprising: scaling the event flux from GW170817 to distances of the other events puts it beyond the sensitivity of most satellites. Indeed, if the gamma ray emission from GW170817 was just 30\% fainter, it would have been missed by current satellites~\citep[see for instance][]{Goldstein2017}. The EM sensitivity to such events is already lagging behind the GW sensitivity to such mergers, despite the latter field being only a few years old. \daksha\ attempts to bridge this gap by drastically improving the sensitivity for high energy transients over current missions.

A closely related field is the study of long and short gamma ray bursts. Great progress has been made in this field, led by missions like BATSE, Swift~\citep{gcg+04} and Fermi~\citep{Meegan2009}. More recently,  \asat\ CZTI~\citep{czti} and POLAR~\citep{2005NIMPA.550..616P} have characterized the polarization of GRBs. 
\daksha\ will detect a large number of gamma ray bursts per year, making significant strides in studies of soft prompt emission, high redshift GRBs, and GRB polarization.
In addition two these two primary science goals, \daksha\ data will give insights into a wide variety of science: including high energy emission in FRBs, studies of accreting pulsars, Soft Gamma Repeaters, monitoring bright sources with Earth Occultation, solar studies, Terrestrial Gamma-ray Flashes (TGFs), and probing primordial black holes with gravitational lensing. In this paper, we describe the design and capabilities of \daksha, and defer a detailed discussion of the science case to Bhalerao et al., (2022) (hereafter \dakshasci).

\section{Daksha}
\begin{figure}
    \centering
     \includegraphics[width=0.6\textwidth]{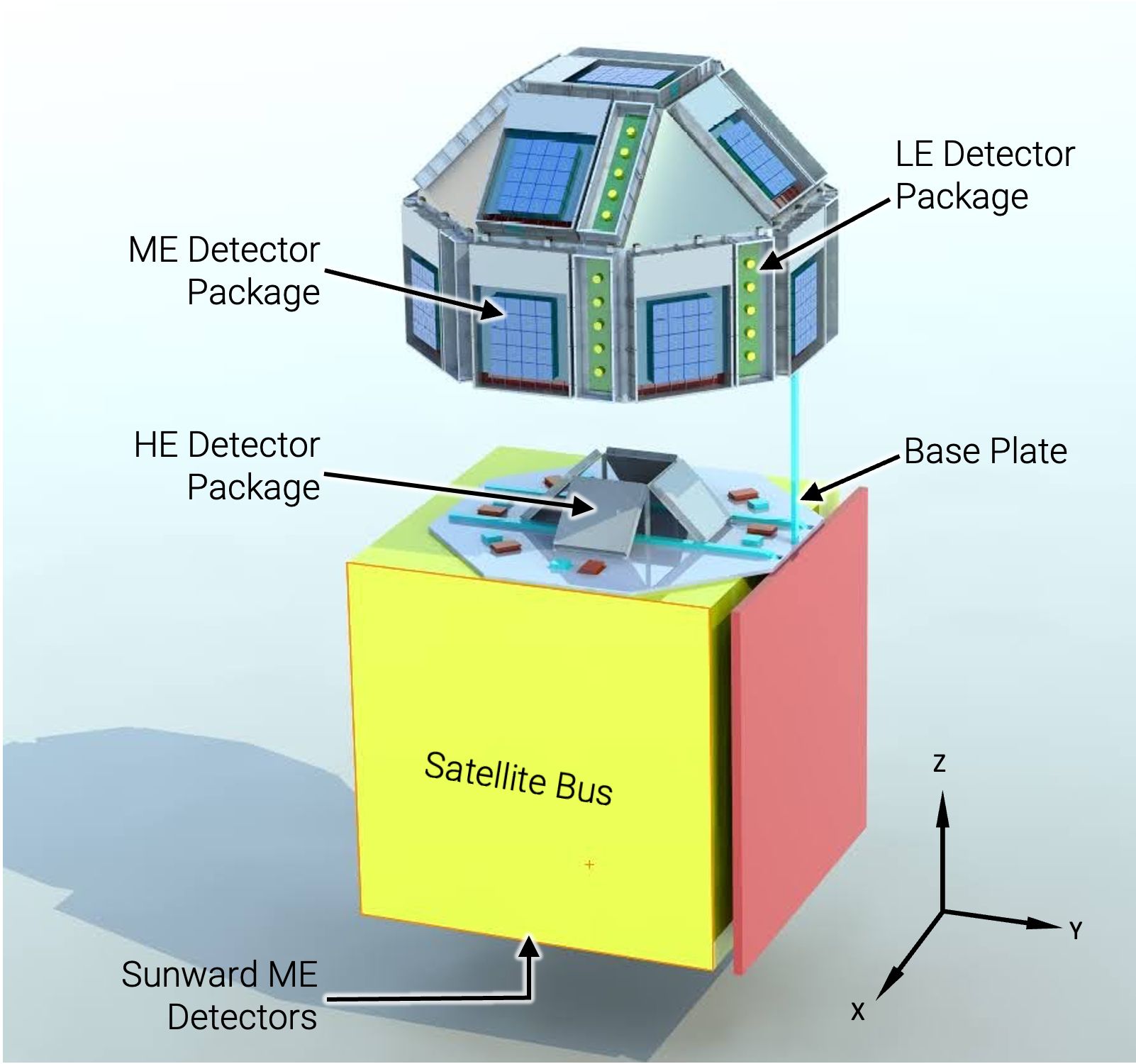}
    \caption{Overall design of a \daksha\ satellite. The dome-shaped payload has 13 surfaces, each carrying Low-energy (LE) and Medium-energy (ME) detector packages. Four ME detector packages are mounted under the satellite bus, and always point directly to the sun. Four High-energy (HE) detector packages are mounted inside the dome, along with processing electronics.}
    \label{fig:config}
\end{figure}
The science goals for \daksha\ can be mapped to three key technical requirements: high sensitivity, continuous all-sky coverage, broadband response. In order to achieve these goals, \daksha\ has been designed as two identical satellites in low-earth orbit (LEO), on opposite sides of the Earth. We aim for an \asat-like near-equatorial orbit with an inclination of 6\degr\ and an altitude of $\sim650$~km~\citep{astrosat}. The broadband response will be attained by using three types of detectors: Silicon Drift Detectors (SDDs), Cadmium Zinc Telluride detectors (CZT), and NaI (Tl) scintillators for Low, Medium and High energies respectively. These are discussed in more detail in Section~\ref{specrange}.

The main payload shape is a partial rhombicuboctahedron\footnote{A rhombicuboctahedron is a solid with eighteen square and eight equilateral triangular surfaces.} with 13 square and 4 triangular surfaces, approximating a hemisphere as shown in Fig~\ref{fig:config}. Medium and low energy detector packages are mounted on each square surface. The ``base plate'' is a flat octagon that carries the high energy detectors, and will be mounted on the satellite bus. Four medium energy detector packages are mounted under the bus.

The satellite will be oriented such that the dome-shaped payload always points away from the sun, and the four medium energy detector packages under the satellite bus will always point at the sun. We refer to the ``anti-sun'' direction (upwards in Fig~\ref{fig:config}) as the boresight. The symmetrical design of the payload ensures good azimuthal uniformity (around the boresight) in the effective area (Section~\ref{allsky}).

\section{Spectral range} \label{specrange}
\daksha\ will attain broadband spectral sensitivity from 1~keV to $>$1~MeV by using SDDs for the low energy band (LE; 1--30~keV), CZT detectors for the medium energy band (ME; 20--200 keV), and NaI Scintillator with Silicon Photomultipliers (NaI + SiPM) for the high energy band (HE; 100--1000+ keV). 
Detectors were selected on the basis of their characteristics like the active area, noise properties, sensitivity, and energy range. All these detectors have been validated for space use in environmental tests or in previous space missions.      

\subsection{Medium Energy}\label{sec:ME}

The workhorses for \daksha\ are sensitive pixelated CZT detectors. We use 5~mm thick detectors manufactured by GE Medical, sensitive to an energy range from 20~keV to 200~keV with an energy resolution of 12\% at 60~keV. Each 3.9~cm $\times$ 3.9~cm detector is divided into a 16$\times$16 pixel array, and integrated electronics provide a digital readout. Detector operating temperatures will be attained using passive cooling with a radiator plate. These detectors have been used before in several space experiments including RT2 \citep{RT2}, High Energy X-ray Spectrometer on Chandrayaan-1 \citep{vps+09}, and CZT Imager on \asat\ \citep{ARRao,czti} where they have proven to be space-worthy and highly effective for this energy range.

\begin{figure}[tbp]
    \centering
    \includegraphics[width=0.75\textwidth]{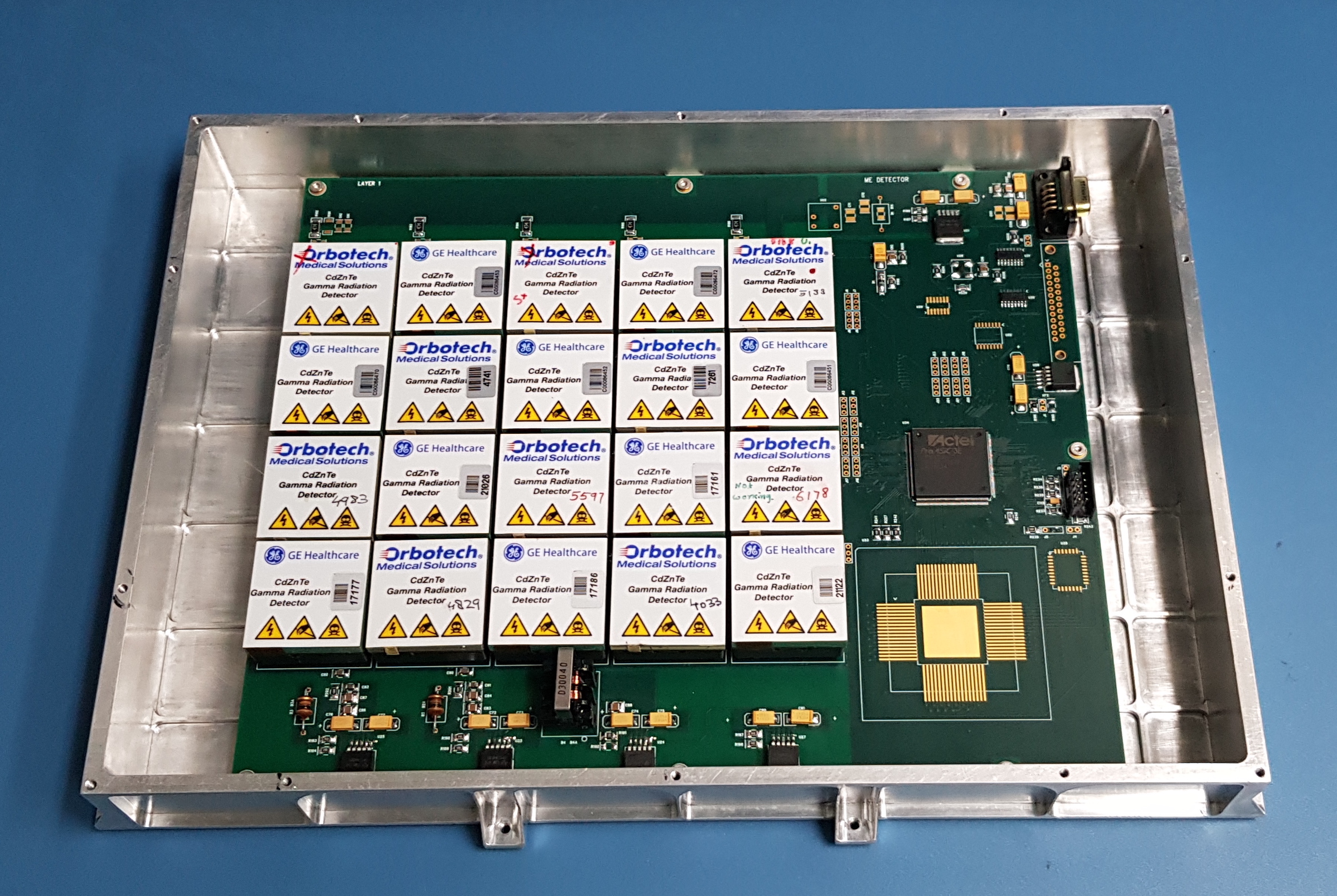}
    \caption{Laboratory model of a \daksha\ ME Package, with all detectors mounted. External connections are not shown in this photograph. Each \daksha\ will have 17 such MEPs.}
    \label{fig:mebox}
\end{figure}

Twenty CZT detectors will assembled into a Medium Energy Package (MEP; Figure~\ref{fig:mebox}) that contains necessary electronics and a heat sink to cool the detectors. The number of detectors in each package determines the sensitivity of \daksha, and is discussed in Section~\ref{s:sens}. Thirteen such identical packages will be placed on the thirteen square surfaces on the dome of \daksha. 
Hence, four MEPs will be mounted on the sunward side of the satellite in order to complete the coverage of the entire sky. This number is chosen to obtain similar effective area as the dome-shaped payload. While the sun is very bright in soft X-rays, solar flux is considerably lower in the ME detector energy range~\citep{Hannah_2010} --- thus these detectors do not risk being saturated in regular observations.

Photon events will be time-tagged with $\sim$microsecond accuracy. Compton scattering can create double events within the detector, which can be used to study polarization of sources as has been demonstrated with \asat\ CZTI \citep{2022ApJ...936...12C,2019ApJ...884..123C,2018NatAs...2...50V,vcr+15}. We have undertaken detailed simulations of the polarization sensitivity, and the results will be presented in Bala et al., in prep. 

\subsection{Low Energy}\label{sec:LE}
\daksha\ Low Energy (LE) detectors have to provide sensitivity down to 1~keV, while overlapping with the ME detector energy range at the higher side. In our baseline design, the low energy coverage will be provided by SDDs with an active thickness of 450~$\mu$m to cover the low energy range from 1--30~keV. We use SDD modules with 1.5~cm$^2$ collecting area, and a thin Beryllium window to block lower energy photons. Low electronic noise is achieved by means of a built-in thermo-electric cooler. These devices have an energy resolution of $\sim$2.5\% (at 5.9 keV) and a timing resolution of 10~$\mu$s. These modules
have been used in the APXS and XSM payloads on Chandrayaan-2 \citep{chandra2,xsm}.

A standard Low Energy Package (LEP) will consist of five Low Energy SDDs, their electronics, and active cooling systems. One LEP will be mounted next to the MEP on each of the thirteen surfaces on the top dome. As the bright sun can saturate LE detectors, no LEPs are mounted on the sunward side of the satellite.

\subsection{High Energy}\label{sec:HE}
At high energies, the coverage of CZT detectors is complemented by Thallium activated Sodium Iodide scintillators, NaI (Tl). Each high energy detector package (HEP) consists of a 20~cm $\times$ 20~cm $\times$ 2~cm NaI(Tl) crystal, sensitive to photons from $\sim$100~keV to $>$1~MeV. The detector works on a sparse array of Si-PM with uniform spatial sensitivity operating on low voltage ($\sim$40~V). This arrangement ensures uniform light collection, and also gives high position sensitivity with spatial resolution of $\sim$1~cm. Photons will be time-tagged with microsecond accuracy. The dome, covered with LE and ME packages, gets progressively transparent at energies $\gtrsim 150$~keV. Hence, we place the HE scintillators inside the dome on the base plate. We opt for four HE detector packages mounted parallel to the four inclined square faces of the dome. This enables us to detect Compton pairs of photons scattered from one of the ME detectors and absorbed in the HE detector, which will help in localizing transient sources and creating an all-sky Compton map of persistent sources.

All the power electronics and processing electronics will be mounted on the base plate.

\section{All-sky coverage}\label{allsky}
The quasi-hemispherical design of the payload naturally results in nearly uniform sensitivity over half of the sky. This design also gives good angular resolution for source localization, discussed in Section~\ref{s:localisation}. The MEPs mounted under the spacecraft bus extend this sensitivity to the rest of the sky as well, though localization abilities become poorer for sources closer to the satellite $-Z$ axis. For a satellite in low-earth orbit at an altitude of $\sim650$~km, about 29\% of the sky is occulted by the Earth. Further, the satellite will spend an average of 24\% of its time in the South Atlantic Anomaly (SAA) region where the high charged particle concentration poses a risk to high voltage electronics and detectors have to be deactivated. As a result, the time-averaged sky coverage of a single satellite in equatorial low earth orbit is limited to about 56\%.

To overcome this limitation, the \daksha\ mission will will deploy two satellites on opposite sides of the Earth. In this configuration we get 100\% sky coverage when neither of the satellites is in the SAA. When either one of the satellites is in SAA, we get 71\% coverage as before. Hence, the total sky coverage of the twin-satellite \daksha\ mission is $\sim 86\%$.

\section{Sensitivity}\label{s:sens}
\begin{figure}[thbp]
    \centering
    \includegraphics[width=0.75\textwidth]{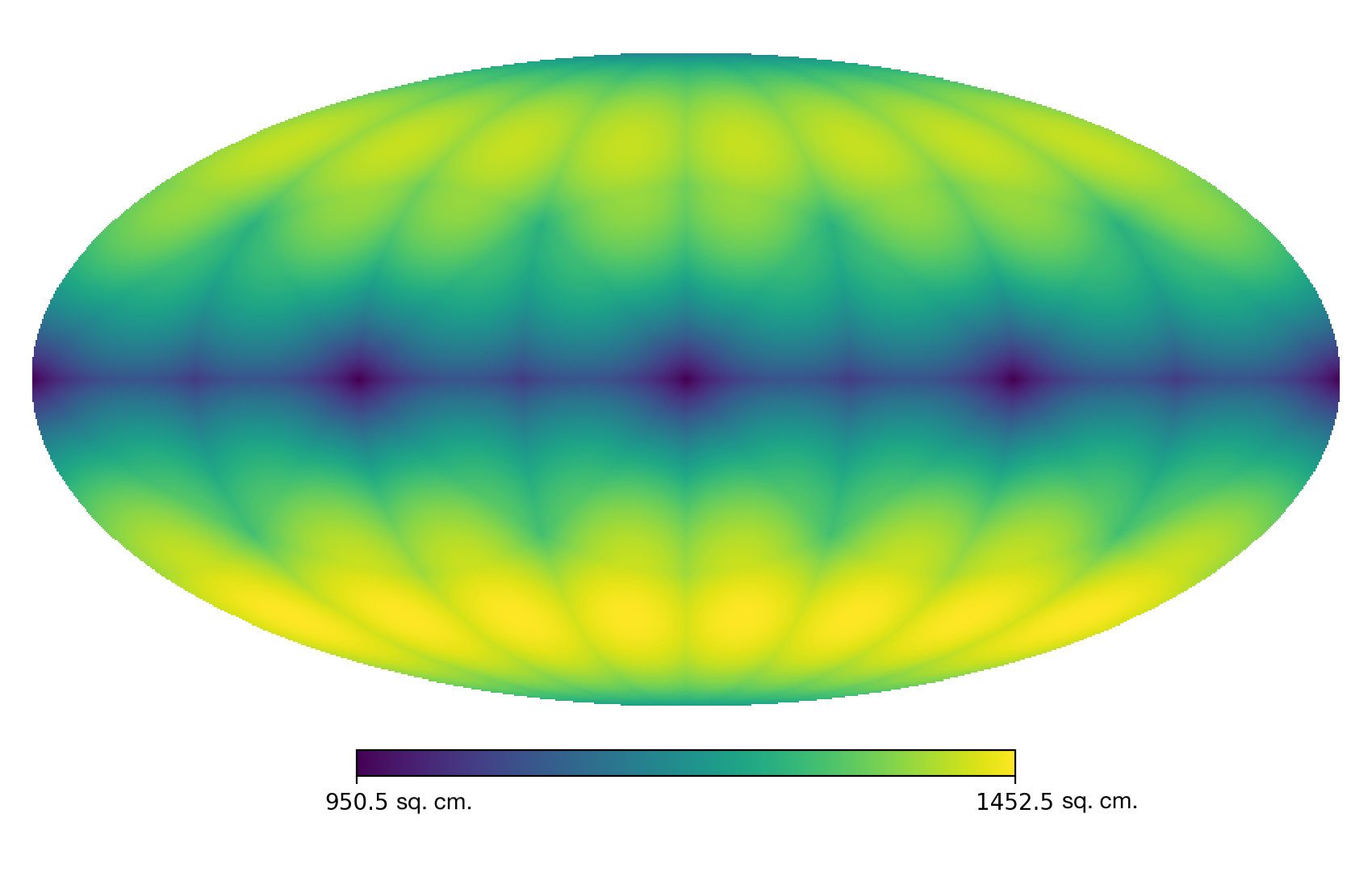}
    \includegraphics[width=0.75\textwidth]{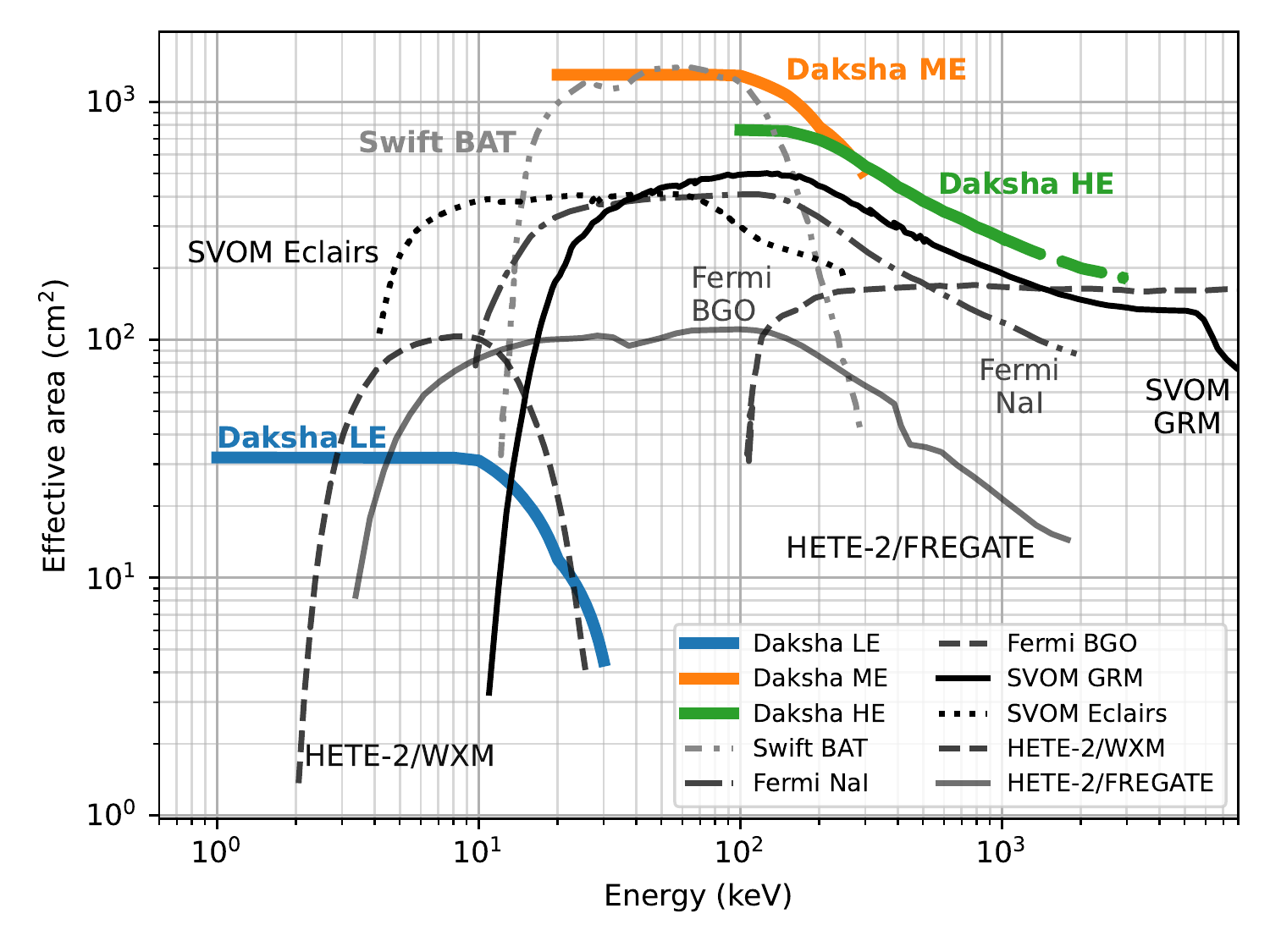}
    \caption{\textit{Top:} Effective area as a function of position, in satellite coordinates. The satellite boresight is at the top, and the sun will always be at the bottom. About 29\% of the sky will be occulted by the Earth at any given instant. \textit{Bottom:} Effective area with respect to detector's energy coverage.  Effective areas of \bat\ (imaging effective area, with mask) and both NaI and BGO detectors of \gbm\ are shown for comparison. These areas were adapted from Luo et al.\citep{2020JHEAp..27....1L} and Sakamoto et al.\citep{Sakamoto2011}. The NaI effective area is the averaged over the unocculted sky.}
    \label{fig:effectArea}
\end{figure}

A key requirement for \daksha\ is to have significantly higher sensitivity for short bursts, as compared to existing missions. The on-axis effective area of each MEP is 304~cm$^2$ on each surface. Considering projection effects over the entire visible sky, the median effective area for \daksha\ ME detectors is 1316~cm$^2$ at 100~keV (ignoring effects of Earth occultation; Figure~\ref{fig:effectArea}). Accounting for the varying satellite-Earth vector through the orbit, the median un-occulted effective area is 1310~cm$^2$. This is significantly higher than the 126~cm$^2$ on-axis effective area for individual NaI modules on {\em Fermi} GBM \citep{Meegan2009}, which has a similar orbit and also covers the entire non-occulted sky. Our effective area is comparable to the masked effective area of the Burst Alert Telescope (BAT) on the Neil Gehrels \swift\ Observatory \citep{gcg+04,Barthelmy2005}, though \bat\ can observe only about 11\% of the sky as opposed to 71\% for each \daksha\ satellite. The effective area gradually tapers off above 100~keV.

A transient located anywhere in the sky will be incident on 4-10 ME packages. Out of these, several will have glancing incidence angles such that the projected area is very small. Ignoring the MEPs with projected area $<40~\mathrm{cm}^2$, the median number of packages that detect any source is 7. In practice, such a cut does not drastically affect the total effective area: after applying this filter, the all-sky median effective area becomes 1296~cm$^2$. To calculate the sensitivity, we assume that the background count rate is same for all ME packages. Thus, we have a source signal corresponding to the median effective area ($\sim 1300~\mathrm{cm}^2$, and background noise corresponding to surfaces on which the transient is incident (typically 7).

The HEPs give a smooth continuation of the MEP effective area at higher energies, extending beyond 1~MeV. The median all-sky effective area for HEPs is 762~cm$^2$. This number will decrease when a detailed model of the satellite bus is included in the calculation. 

The LEPs have a median effective area of 32.0~cm$^2$ in the anti-sun direction, accounting for projection effects. The effective area slowly drops off in the sunward direction, and the all-sky median effective area is 26.0~cm$^2$.

The flux sensitivity depends on the source spectrum and background rates. We estimate the background with a GEANT4 simulation that includes the cosmic diffuse X-ray background, particle-induced backgrounds, and albedo components. We also add a component for electronic noise, estimated from the CZTI instrument on \asat\ \citep{czti}. By considering all components, we estimate the net background count rate to be about $10~\mathrm{counts~cm}^2~\mathrm{s}^{-1}$.

\begin{sidewaystable}
\begin{center}
\caption{\centering Comparison of key parameters of GRB missions.\label{tab:comparemissions}}
\footnotesize
\begin{tabular}{|l|c|c|c|c|c|c|c|c|c|}
\toprule
\multirow{2}{*}{Mission name} & Energy range & Effective area & \multicolumn{2}{c|}{FoV} & Range & Volume & \multicolumn{2}{c|}{Sensitivity (1-s, 5$\sigma$)} & Reference\\
\cline{4-5}
\cline{8-9}
& keV & (cm$^2$) & Sky fraction & (sr) & Mpc & Mpc$^3$ & \ecs & \pcs &\\
\midrule
\daksha\  (single) & 20--200 & 1300 & 0.7 & 8.8   & 76 & 1.27\ee{6}   & 4\ee{-8} & 0.6 & This work\\
\daksha\  (two) & 20--200 & 1700 & 1 & 12.6       & 76 & 1.81\ee{6}   & 4\ee{-8} & 0.6 & This work\\
\bat & 15--150 & 1400 & 0.11 & 1.4                & 67 & 0.14\ee{6}   & 3\ee{-8} & 0.5 & \cite{Lien2014}\footnotemark[1]\\
\gbm & 50--300 & 420 & 0.7 & 8.8                  &  49 & 0.35\ee{6}   & 20\ee{-8} & 0.5 & \cite{2020ApJ...893...46V, 2021ApJ...913...60P}\footnotemark[2] \\
GECAM-B & 6--5000 & 480 & 0.7 & 8.8               &  65 & 0.81\ee{6}   & 9\ee{-8} & --- & \cite{gecam_poster,Zhang2019}\\
\eclairs & 4--150 & 400 & 0.16 & 2                & 70 & 0.23\ee{6}   & 4\ee{-8} & 0.8 & \cite{Lacombe2019}\\
\xgis & 2--30 & 500 & 0.16 & 2                    &  45 & 0.06\ee{6}   & 1.7\ee{-8} & --- & \multirow{3}{*}{\cite{2021arXiv210208701L}} \\
\xgis & 30--150 & 500 & 0.16 & 2                  &  58 & 0.12\ee{6}   & 5\ee{-8} & --- & \\
\xgis & 150--1000 & 1000 & 0.5 & 6.2              &  20 & 0.02\ee{6}   & 45\ee{-8} & --- & \\
\botrule
\end{tabular}
\footnotetext{Sensitivity numbers are scaled to 5-$\sigma$ threshold for a 1~s transient. The sensitivity numbers should be compared with caution: for instance for power-law spectrum with photon index $\alpha = -1$, the \bat\ 15--150~keV sensitivity limit of 3\eu{-8}{\ecs} is equivalent to the  \daksha\ 20--200~keV sensitivity limit 4\eu{-8}{\ecs}. The Field of View is the peak FoV, and does not include any temporal downtime corrections. \\
``Range'' and ``Volume'' are calculated for GW170817 as discussed in the text. \bat\ range is taken from Tohuvavohu et al.\cite{2020ApJ...900...35T}, our estimate was higher (79~Mpc), the difference likely comes from not accounting for reduced number of detectors currently operational. \gbm\ calculated range is consistent with~\cite{2020ApJ...900...35T,Goldstein2017}. Range for other instruments is calculated from the sensitivity numbers used in this table.\\
The sky coverage is instantaneous. This table does not correct for any instrument downtime from SAA passages or any other causes.}
\footnotetext[1]{\url{https://swift.gsfc.nasa.gov/about_swift/bat_desc.html}}
\footnotetext[2]{\url{https://fermi.gsfc.nasa.gov/science/instruments/table1-2.html}}
\end{center}
\end{sidewaystable}

To be conservative and avoid false negatives, we select a detection threshold as a signal-to-noise ratio of 5, for a burst of 1 second duration. Assuming a typical GRB-like Band spectrum, the faintest sources that \daksha\ can detect have a flux of $4\times 10^{-8}~\ecs$ (1~s, 5-$\sigma$). The numbers of course depend on burst duration and the sensitivity threshold. For instance, the 3-$\sigma$ limit for a 20~s burst is 0.5\ee{-8}{~\ecs}. Adopting a more conservative threshold for short bursts, the 7-$\sigma$ sensitivity for 1~s bursts is 5.6\ee{-8}{~\ecs}. 

In Table~\ref{tab:comparemissions}, we compare the sensitivity of \daksha\ with various other missions, scaled to the same detection criteria. We calculate the distance to which the missions would be sensitive to GW170817, assuming a ``Comptonized'' spectral model with a photon power-law index $\alpha = -0.62$, an exponential cut-off at $E_p = 185$~keV, duration $\Delta t = 0.576$~s \citep{Goldstein2017}. The luminosity is calculated to match the observed Fermi flux of 3.1\eu{-7}{~\ecs} in the 10--1000~keV band.
The ``range'' column of the table shows the distance to which such a source could be detected by each mission. The relevant figure of merit is now the total volume probed by each mission, derived from the range and the field of view. We can see that \daksha\ probes a significantly higher volume than any other mission.

\section{Localization}\label{s:localisation}
The focus of the \daksha\ mission is to detect transients, and we trade off angular resolution for all-sky coverage. Based on our calculated background rates, we find that a short GRB ($T_{90}=1$~s) with a fluence of $10^{-7}~\mathrm{erg~cm}^{-2}$ can be localized with an accuracy of 10\degr. The localization accuracy depends on the source spectrum, the $T_{90}$, and also on the position of the transient in the satellite frame ---  with worse angular resolution on sunward side of the satellite. For a given fluence, a longer GRB will be localized more poorly due to the higher number of background counts in that duration. Angular resolution is approximately inversely proportional to flux, such that a 1~s GRB with flux $10^{-6}~\mathrm{erg~cm}^{-2}~\mathrm{s}^{-1}$ can be localized to an accuracy of $\approx 1\degr$. At this flux level, the background contribution becomes comparable to the number of source photons in the ME detectors.

For transients detected by both satellites, the location can be improved by joint triangulation of the source. On-board clocks on \daksha\ will have an absolute timing accuracy of at least 1~ms. For transients with sufficiently high fluence, joint analysis of data from both satellites (including projection effects and light-travel delay) can be used to obtain source positions with 1--2\degr\ accuracy. \daksha\ could also be used along with other satellites analogous to the Interplanetary Network (IPN\footnote{\url{http://www.ssl.berkeley.edu/ipn3/}}), to obtain more precise transient locations.

\section{Alerts and data}
\daksha\ satellites will have robust on-board processing capabilities to detect transients. For each detected transient, on-board software will calculate the source position, spectrum, and $T_{90}$. It will also create a light curve in multiple bands, and low-resolution spectra to create a $\sim 2$~kB data package. Our mission requirements are to be able to download this information within 1 minute of transient detection, for $>$95\% of the orbit. Transient detections will be announced immediately through standard channels like the General Coordinates Network\footnote{\url{https://gcn.gsfc.nasa.gov/}, formerly GRB Coordinates Network.}~\citep{Barthelmy2003,2022GCN.32419....1B}.

All time-tagged event data from all detectors will be stored on-board, and will be downlinked on every ground station pass. Offline analysis will account for secondary effects like scattering of photons within the satellite and the Earth albedo, to produce more accurate localizations and spectra. All data and data products will be made public after a limited proprietary period, using standard FTOOLs-compatible FITS formats~\citep{1981A&AS...44..363W,blackburn95}.

\section{International Scenario}

The \daksha\ mission has excellent synergy with the rapidly growing field of time--domain astrophysics. \daksha\ will search for counterparts to events from the advanced gravitational-wave detector network of  LIGO \citep{aLIGO}, Virgo \citep{AdVirgo}, KAGRA \citep{kagra}, and LIGO-India \citep{indigo,2022CQGra..39b5004S}. \daksha\ transients can be triggers to optical synoptic and survey telescopes like the Vera Rubin Observatory \citep{laa+09} and the Zwicky Transient Facility \citep[ZTF;][]{bellm14}. Conversely, \daksha\ data can also be searched for high energy bursts associated with rapidly evolving transients discovered in such surveys as has often been done with \asat\ and other missions \citep[See for instance][]{bkb+17,2021ApJ...918...63A}. Similar synergy will also exist with radio facilities like the Square Kilometer Array \citep{2009IEEEP..97.1482D}, Australian Square Kilometer Array Pathfinder \citep{2007PASA...24..174J}, Low-Frequency Array \citep[LOFAR;][]{2013A&A...556A...2V}, upgraded Giant Metrewave Radio Telescope \citep[uGMRT;][]{2017CSci..113..707G}, etc.

The current key spacecraft for high energy transient studies --- the Neil Gehrels Swift Observatory \citep{gcg+04} and \fermi\ \citep{Meegan2009} --- are over fifteen years old. Indeed, the Astro2020 decadal survey concluded that, ``In space, the highest-priority sustaining activity is a space-based time-domain and multi-messenger program of small and medium-scale missions'' \citep{astro2020}.
A recent mission in the field is {\em GECAM} \citep[Gravitational wave high-energy Electromagnetic Counterpart All-sky Monitor;][]{Zhang2019}, launched in late 2020. While the mission concept is very similar to \daksha, {\em GECAM} is based completely on scintillator-based detectors, and has lower sensitivity \citep{gecam_poster}. Another upcoming high energy transient mission is {\em SVOM} \citep[Space Variable Objects Monitor;][]{Cordier2015} is a multi-wavelength mission with capabilities to detect GRBs. The CZT-based ECLAIRS instrument \citep{Godet2014} is a coded mask imager with field of view of $\sim2$~steradians, and active area of 1024~cm$^2$ (40\% mask transparency). It is complemented at higher energies by GRM, covering the same part of the sky with sensitivity similar to {\em Fermi}-GBM \cite{Wei2016}. Several teams have also proposed smaller satellites like BurstCube~\citep{2017arXiv170809292R}, HERMES~\citep{2019NIMPA.936..199F}, BlackCAT~\citep{Chattopadhyay2018} etc, with limited effective areas. With significantly higher sensitivity and grasp compared to all these, \daksha\ will revolutionize our understanding of highly energetic transients.

\section{Summary and Current Status}
\daksha\ is an ambitious high energy transients mission designed to have high sensitivity, all-sky coverage, and broadband spectral response. The science of \daksha\ is discussed in detail in \dakshasci, we only list the highlights here. \daksha\ will detect high-energy emission from about 3--12 off--axis neutron star bursts each year, depending on assumed model. In addition, \daksha\ can discover up to 7 more each year which are just sub-threshold for the GW detector networks, but maybe recovered given the confident electromagnetic trigger.
\daksha\ will detect about 500 long GRBs and 50 short GRBs each year and provide time-tagged event data from 1~keV to $>1$~MeV for each of these events. \daksha\ will be the first all-sky monitor to obtain soft X-ray spectra of these transients. Furthermore, \daksha\ will address a wide range of science cases including studying GRB hard X-ray polarisation, monitoring X-ray pulsars, studies of magnetars, solar flares, searches for fast radio burst counterparts, routine monitoring of bright persistent high energy sources, terrestrial gamma-ray flashes, and probing primordial black hole abundances through lensing.

\daksha\ was proposed in response to the Indian Space Research Organisation's Announcement of Opportunity (AO) for Astronomy missions in 2018. The proposal was shortlisted for further studies, and was awarded seed funding by the Space Science Program Office for demonstration of a proof-of-concept. The team has completed the construction and testing of a laboratory model of the Medium Energy detector Package, as required. \daksha\ builds heavily on the legacy of various Indian space science missions, giving a high technology readiness level to all subsystems. The mission will be reviewed for full approval, after which we target a development timeline of three years to launch.

\section*{Acknowledgments}
We thank the Space Program Office (SPO) of the Indian Space Research Organisation for its Announcement of Opportunity for space astrophysics missions, under which \daksha\ was proposed. Development of the \daksha\ MEP laboratory model was started with funding support from SPO. We thank the administrative and support staff at all partner institutes for their help in all \daksha-related matters.

DS was supported under the CEFIPRA Grant No. IFC/5404-1. AP acknowledges the SERB Matrics grant MTR/2019/001096 and SERB-Power-fellowship grant SPF/2021/000036 of Department of Science and Technology, India for support.

\subsection*{Author Contributions}
Varun Bhalerao, Santosh Vadawale, Shriharsh Tendulkar, Dipankar Bhattacharya, Vikram Rana, Gulab Dewangan, Archana Pai, Disha Sawant contributed to the overall mission concept and design. Sujay Mate, Akshat Singhal, Gaurav Waratkar, Divita Saraogi, Sourav Palit contributed to design optimization and simulations. Prabhu Ramachandran contributed to software, simulations, and training.
Hrishikesh Belatikar, Mahesh Bhaganagare, Abhijeet Ghodgaonkar, Jayprakash G. Koyande, Sanjoli Narang, Ayush Nema, Sudhanshu Nimbalkar, Amit Shetye, Siddharth Tallur, Sandeep Vishwakarma, Piyush Sharma, Shiv Kumar Goyal, N.P.S. Mithun, C. S. Vaishnava, Nishant Singh, M. Shanmugam, B.S. Bharath Saiguhan, Arpit Patel, Tinkal Ladiya, Priya Pradeep, S. Sreekumar, Suddhasatta Mahapatra contributed to electronics development.
Hitesh Kumar L. Adalja, Neeraj K. Tiwari, APK Kutty, Suresh Gunasekaran, Guruprasad P J, Salil Kulkarni, Rakesh Mote, Srividhya Sridhar contributed to structural design and simulations. Deepak Marla, Gaurav Waratkar, Shreeya Singh, Amrutha Lakshmi Vadladi contributed to thermal design and simulations. Jinaykumar Patel, Rahul Srinivasan contributed to the orbital simulator.

\subsection*{Conflict of Interest}
The authors declare that there are no conflicts of interest in this manuscript.

\subsection*{Funding}
Development of Daksha was supported by a seed grant awarded to IIT Bombay from the Indian Space Research Organization. Further expenses were also supported by the partner institutes including direct and in-kind contributions.

\subsection*{Software}
Numpy~\citep{numpy}, Matplotlib~\citep{matplotlib}, Astropy \citep[\url{http://www.astropy.org}]{astropy}, 
  HealPIX~\citep{ghb+05}, Healpy (\url{https://healpy.readthedocs.org/}), Ephem (\url{https://pypi.python.org/pypi/pyephem/}).

\bibliography{daksha}


\end{document}